\documentclass[a4paper,12pt]{iopart} 
\usepackage[latin1]{inputenc}

\usepackage{amssymb,bbm,epsfig,ulem,dsfont,color}

\newcommand{\be}{\begin{equation}}
\newcommand{\ee}{\end{equation}}
\newcommand{\ba}{\begin{array}}
\newcommand{\ea}{\end{array}}
\newcommand{\bqa}{\begin{eqnarray}}
\newcommand{\eqa}{\end{eqnarray}}

\newcommand{\ie}{{\it i.e.}}
\newcommand{\eg}{{\it e.g.}}

\newcommand{\bra}[1]{\left\langle{#1}\right|}
\newcommand{\ket}[1]{\left|{#1}\right\rangle}

\newcommand{\fig}[1]{Fig.\,\ref{#1}}
\newcommand{\eq}[1]{Eq.~(\ref{#1})}

\begin{document}
\normalem

\title{Robust states of ultra-cold bosons in tilted optical lattices}

\author[M. Hiller et al.]{Moritz Hiller$^{1,2}$, Hannah Venzl$^1$, Tobias Zech$^1$, Bart\l{}omiej Ole\'{s}$^3$, Florian Mintert$^{1,4}$, and Andreas Buchleitner$^1$}

\address{$^1$Physikalisches Institut, Albert-Ludwigs-Universit\"at Freiburg, Hermann-Herder-Str. 3, 79104 Freiburg, Germany}

\address{$^2$Institute for Theoretical Physics, Vienna University of Technology, Wiedner Hauptstra{\ss}e 8-10/136, A-1040 Vienna, Austria}

\address{$^3$Marian Smoluchowski Institute of Physics and Mark Kac Complex Systems Research Center,
Jagiellonian University, Reymonta 4, 30-059 Krak\'ow, Poland}

\address{$^4$FRIAS, Freiburg Institute for Advances Studies, Albert-Ludwigs-Universit\"at Freiburg, Albertstr. 19, 79104 Freiburg, Germany}


\begin{abstract}
We identify regular structures in the globally chaotic spectra of an interacting bosonic quantum gas in tilted periodic potentials.
The associated eigenstates exhibit strong localization properties on the lattice, and are dynamically robust against external perturbations.
\end{abstract}

\maketitle

\section{Introduction}
Ultra-cold atoms in optical lattices provide a versatile testing ground for the physics of interacting many-body quantum systems,
ranging from the characteristic properties of many-particle ground states \cite{DGPS99}, over atomic transport \cite{Fert05,PMKB06}, to
the emergence of collective properties, and thermodynamic behavior \cite{MO06,BDZ08}.
The complexity of the many-body dynamics of such systems increases rapidly with the number of particles and lattice sites
and has its characteristic spectral counterpart in a highly irregular parametric evolution of the energy levels \cite{FFS89,Chef96,KB04,HKG09}.
This implies a high sensitivity of the time evolution with respect to changes in the initial conditions and/or perturbations
of the generating Hamiltonian, and renders control of generic many-particle dynamics an extremely challenging task.
However, nonlinear coupling can also give rise to the emergence of stable collective modes which opens new perspectives
for robust control \cite{Chirikov79,Buchleitner2002409}.
In our present contribution, we will identify such modes for the tilted Bose-Hubbard Hamiltonian (BHH), 
and demonstrate their pronounced localization properties as well as their extraordinary robustness against perturbations.
In contrast to previous experimental and theoretical studies on bound-atom states (see, e.g., Refs.~\cite{Wink_etal06,DESS04,VPS10}) the stability of these solutions is not a mere consequence of energy separation (by a spectral gap) due to the presence of interatomic interactions. 
Instead, we find bound states that exist within the bulk of the spectrum which can be considered chaotic in the sense of random matrix theory \cite{Haa00}.

\section{Model}
The simplest quantum mechanical many-body description of ultra-cold bosons in a lattice is the Bose-Hubbard Hamiltonian
\cite{JBCGZ98} that incorporates both, the tunneling of individual particles between neighboring sites,
and their pairwise on-site interaction. When the one-dimensional lattice is subject to an additional static tilt
(due to, \eg, gravitation), the Hamiltonian takes the form
\begin{equation}
\hat H=-\frac{J}{2}\sum_{l=1}^{M-1} (\hat a_{l+1}^\dagger \hat a_l +h.c.)
+\frac{U}{2}\sum_{l=1}^{M} \hat n_l (\hat n_l-1)+F\sum_{l=1}^{M} \tilde{l} \hat n_l \,,
\label{Eh}
\end{equation}
where $\hat a_l$ ($\hat a_l^\dagger$) annihilates (creates) a particle in the Wannier state localized at the $l$-th
site, $\hat n_l=\hat a_l^\dagger\hat a_l$ is the associated number operator, and $M$ specifies the length of the lattice.
Here we consider a tilt around the center of the lattice, and the on-site term $\tilde{l}$ hence takes the values
$\tilde{l}=-M/2+l$ for even and $\tilde{l}=-(M+1)/2+l$ for odd $M$.
The BHH has two constants of motion, the energy $E=\langle\hat H\rangle$ and the total number of particles $N=\langle\sum_l\hat n_l\rangle$.
The parameters $J$, $U$, and $F$ describe the tunneling strength, the on-site interaction and the static tilting field, respectively.

The model is based on a single (lowest) band approximation for the optical lattice \cite{JBCGZ98}.
This assumption is valid provided the kinetic energy, the interaction strength, and the local chemical potential
(resulting from the tilt), are sufficiently small not to excite higher Bloch bands.
Therefore, the lattice needs to be sufficiently deep \cite{JBCGZ98,DGPS99}, to induce large band gaps,
and the interaction energy must be smaller than the single-particle ground-state energy,
in order not to modify the single-particle wave function  considerably.
In the experiment, these conditions can be met since all the parameters $J$, $U$, and $F$ in the BHH are readily controlled \cite{MO06}:
While $J$ and $F$ are solely determined by the lattice geometry, the inter-atomic interaction $U$
can additionally be adjusted using Feshbach resonances \cite{IASMSK98, CGJT10}.

In the limit of large particle numbers  $N\gg1$, the quantum dynamics may be described by the mean-field counterpart of the Bose-Hubbard Hamiltonian, the discrete Gross-Pitaevskii equation (GPE) (see, \eg, \cite{ELS85,SFGS97}).
At fixed lattice length $M$, the mean-field limit is approached by increasing the particle number and, at the same time, keeping the scaled interaction $UN$ at a constant value.
As a mean-field approach, the discrete GPE does not explicitly contain the particle number $N$ and, hence, hardly covers effects related to the granularity of matter
\footnote{
We note that for small particle numbers, deviations of the GPE description from the many-body picture have been predicted, e.g., for a two-site lattice in Refs.~\cite{VA00,AV01}, where a comparison to the Bose-Hubbard dynamics was performed.
In Refs.~\cite{SSAC09,SSAC10}, both the BHH and the continuous GPE were further compared to the {\it multiconfigurational time-dependent Hartree for bosons method}, a more advanced propagation scheme for the many-body Schr\"odinger equation.
The applicability of the GPE can though be extended, when combined with phase-space methods, i.e., when one considers not a single mean-field trajectory but propagates an ensemble of GPEs that reflects the initial quantum mechanical state (see, e.g., \cite{Chuc10} and references therein).
}.
In the following, we are concerned with a rather small number of interacting bosons where such effects will be essential, and thus resort to the full many-body Hamiltonian (\ref{Eh}).

\begin{figure}[t]
\begin{center}
\includegraphics[width=.9\textwidth,type=pdf,ext=.pdf,read=.pdf,keepaspectratio,clip] {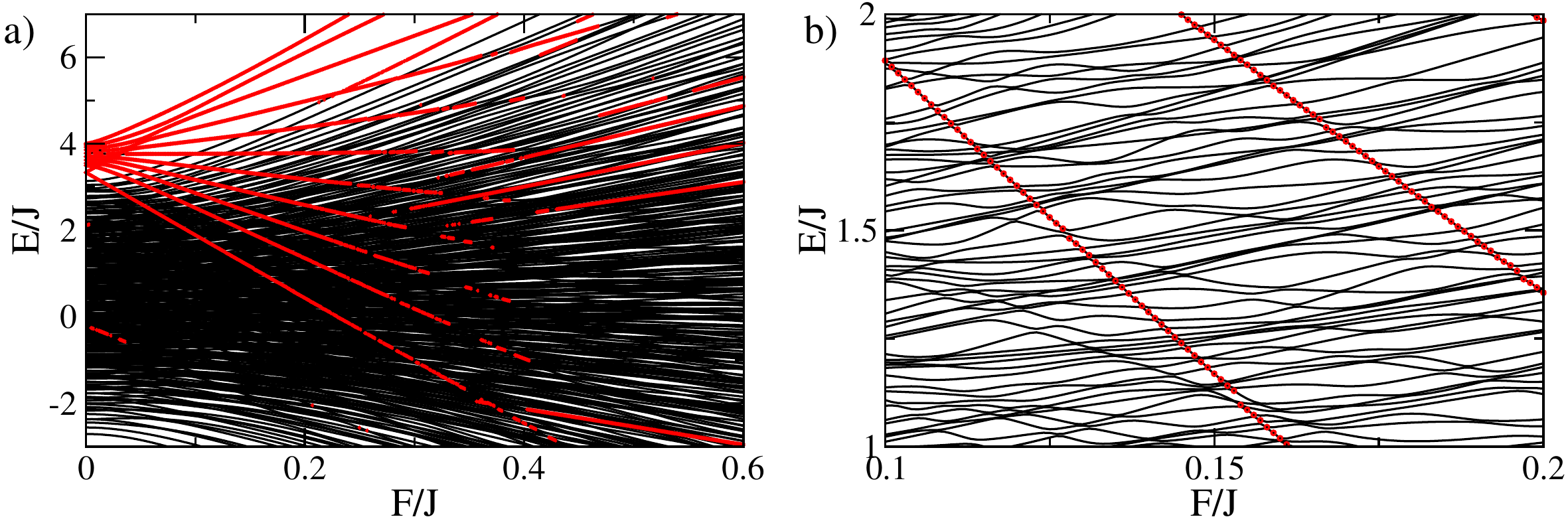}
\caption{
(Color online) Spectrum of the Bose-Hubbard-Hamiltonian (Eq.~(\ref{Eh})) as a function of the static tilt $F/J$, with interaction strength $U=1$ and tunneling coupling $J=1$, for $N=3$ particles and $M=11$ lattice sites.
The $M$ eigenstates with the largest IPR (see Eq.~(\ref{Eipr})) in the Fock number basis are plotted in red. a) A set of (almost) straight-line energy levels traverse the chaotic background, without changing the slope.
This set is characterized by the localization of the corresponding {\em solitonic} eigenstates in the Fock
basis. b) A zoom into the vicinity of the two lowest solitonic levels reveals small avoided crossings.
}
\label{Fspectrum}
\end{center}
\end{figure}

\section{Parametric level evolution}
Tunneling, on the one hand, and interaction and tilt, on the other, define two (incompatible) symmetries of the system Hamiltonian (\ref{Eh}).
Hence, if either one of these terms dominates, the many-particle eigenstates exhibit the relatively simple
structure of Bloch waves (for $J\gg UN,F$) or Wannier states (for $UN\gg J,F$ or $F\gg J,UN$).
In the generic case, however, when all three terms have comparable weight, good quantum numbers with an unambiguous
labeling of the system eigenstates cannot be defined, since the energy levels exhibit a complicated parametric evolution with
$U$, $J$, or $F$, and avoided crossings of variable size abound \cite{KB04,HKG09,KB03,Venzl11}.
This is the spectral manifestation of {\em quantum chaos}, and is nicely illustrated by the overwhelming number of
energy levels in \fig{Fspectrum}.

\begin{figure}
\begin{center}
\includegraphics[width=.6\textwidth,type=pdf,ext=.pdf,read=.pdf]{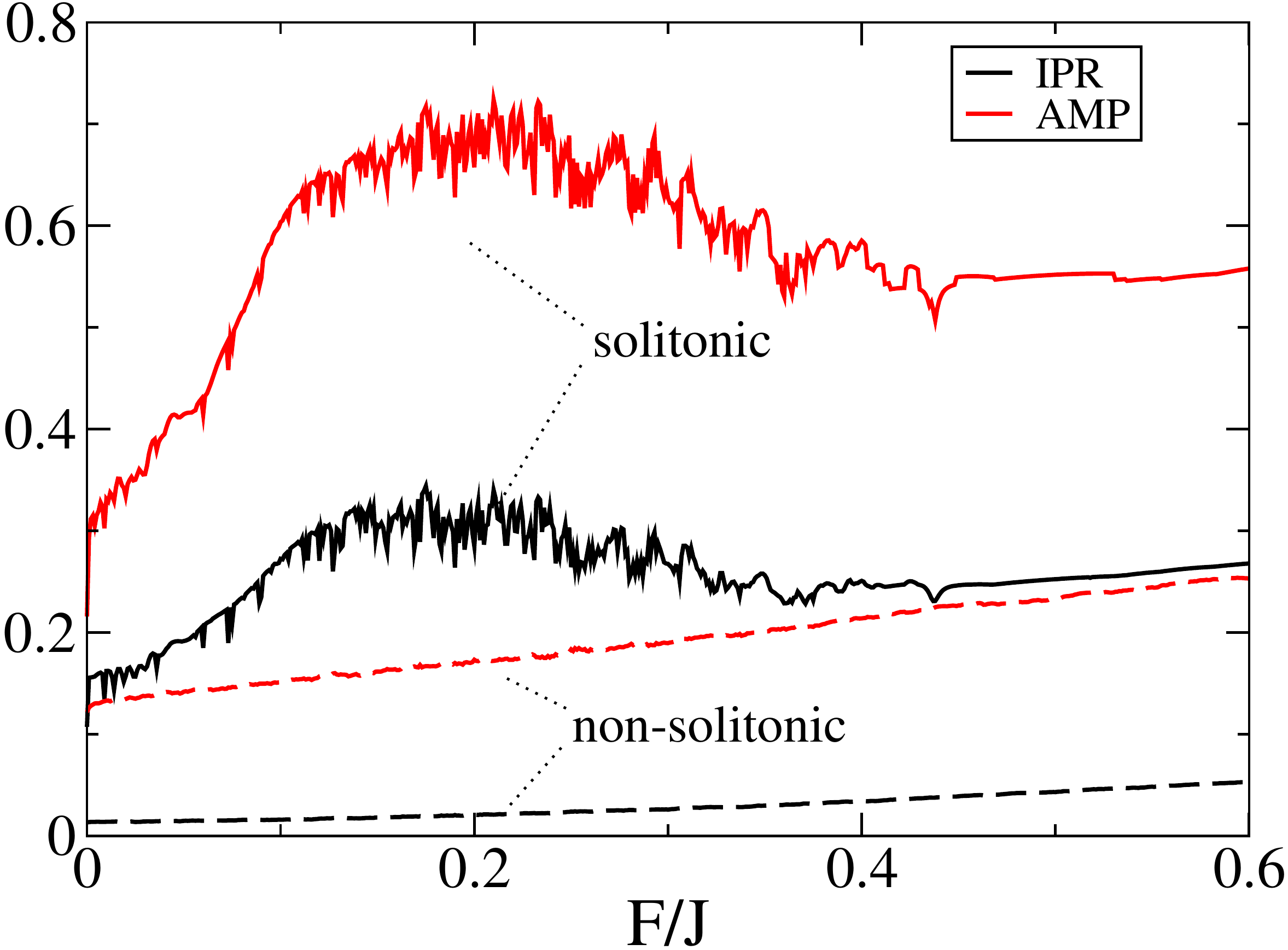}
\caption{ (Color online)
The average maximal population (AMP, red), \eq{amp},  and the averaged inverse participation ratio (IPR, black), \eq{Eipr}, of the eigenstates of the Bose-Hubbard Hamiltonian, \eq{Eh}.
The solid lines correspond to the average of the $M=11$ eigenstates with the largest individual IPR,
the dashed lines to the average of all the other eigenstates.
{Within the range $0.1\lesssim F/J\lesssim 0.4$, the solitonic states are strongly localized, both on the lattice and in the Fock-space.}
The parameters are chosen as in \fig{Fspectrum}, and the dimension of the Hilbert space is ${\cal N}=286$.}
\label{FIPR}
\end{center}
\end{figure}

However, the figure also shows energy levels with constant slope over a wide interval of the static tilt $F$ (at fixed $U$ and $J$),
which thus represent many-particle eigenstates with characteristic properties which are invariant under changes of $F$:
By virtue of the Hellmann-Feynman theorem \cite{Feyn39},
\begin{equation}
\frac{\partial E}{\partial F}=\left\langle\frac{\partial \hat H}{\partial F}\right\rangle=\left\langle \sum_l \tilde{l}\hat n_l\right\rangle\,,
\label{Ehf}
\end{equation}
where $\langle\cdot\rangle$ represents the expectation value with respect to a single such state,
the slope of these energy levels defines a constant center of mass of the many-particle wave function, and, hence,
suggests invariant localization properties of the particles on the lattice, under changes of $F$.
%
%
This is corroborated by the average maximal particle number (AMP) of these states on a single
lattice site
\begin{equation}
  \label{amp}
  {\rm AMP}=\left\langle\frac{\max_l\bra{\psi}\hat n_l\ket{\psi}}{N}\right\rangle \ ,
\end{equation}
in \fig{FIPR}, in comparison to the AMP of the chaotic, \ie, irregular states of the spectrum. The average $\langle\cdot\rangle$
in (\ref{amp}) runs over the respective sample of states, at fixed value of $F$.
Clearly, in the irregular eigenstates the particles are largely spread out over the lattice, with AMP $\lesssim 0.25$, while the solutions with constant slope in \fig{Fspectrum} are stronger localized on the lattice.
Whereas this localization is to some extent present already for vanishing tilt, it becomes strongest once the corresponding energy levels cross the bulk, \ie, for tilt strengths $0.1\lesssim F/J\lesssim 0.4$, where more than $60\%$ of all ($N=3$) particles are localized on a single site.
Because of their pronounced localization properties, that we will show to remain unchanged under variation of the tilt, we refer to these as {\em solitonic} states  in the following \cite{ZBD97}.
In contrast to previously identified localized states \cite{Wink_etal06,DESS04,VPS10}, the presently discussed solitonic states are not merely a consequence of energetic isolation with respect to the remaining part of the spectrum.
Finally, we note that the observed localization is consistent with the fact that there exist as many solitonic states as sites of the lattice --
$M=11$ in the case investigated in Figs.~\ref{Fspectrum} and \ref{FIPR}.

%
Given the fact that the solitonic states are localized around one lattice site, we further check whether they also exhibit localization in Fock space.
A measure for the latter is the inverse participation ratio (IPR) with respect to the Fock basis
$\ket{b_j}=\ket{n_{j,1},n_{j,2},...n_{j,M}}$, which forms the eigenbasis of the BHH for vanishing tunneling $J=0$.
The IPR is defined as:
\begin{equation}
{\rm IPR}(\ket{\psi})=\sum_{j=1}^{{\cal{N}}}|c_j|^4\ ,
\label{Eipr}
\end{equation}
where the $c_j$ are the expansion coefficients of the state $\ket{\psi}$ in a given basis, and ${\cal{N}}$ is the Hilbert-space dimension.
The IPR represents the inverse number of basis states that are occupied by the state $\ket{\psi}$,
and varies from unity -- when $\ket{\psi}$ coincides with one basis state -- to $1/\cal{N}$, for $\ket{\psi}$ an
equally-weighted superposition of all basis states.
In fact, the states highlighted in red in \fig{Fspectrum} are precisely those eleven states with the largest IPR, \ie, those which exhibit
the strongest localization properties in Fock space.
To quantify this statement we calculate the IPR averaged over the eleven solitonic states, as well as for all non-solitonic states
\footnote{
The fluctuations of AMP and averaged IPR for the solitonic states, as visible in \fig{FIPR},
can be attributed to avoided crossings that locally affect the levels and are not washed out due to the rather small number
of solitonic levels that enter the average. For the non-solitonic states, AMP and averaged IPR show less fluctuations,
since the average is taken over significantly more states.}.
For the latter, \fig{FIPR} shows a moderate and monotonic increase of the IPR with growing $F$, what can be attributed
to the onset of Stark localization, as expected for a very large tilt (see, e.g. \cite{GGK02}).
{In contrast, the IPR of the solitonic states is about one order of magnitude larger than the IPR of the irregular states,} for $0.15\lesssim F/J\lesssim 0.35$.
For stronger tilt, some of the solitonic states disappear what we discuss in Section {\it Generating mechanism} further down.

\begin{figure}[t]
\begin{center}
\includegraphics[width=.6\textwidth,type=pdf,ext=.pdf,read=.pdf,keepaspectratio,clip] {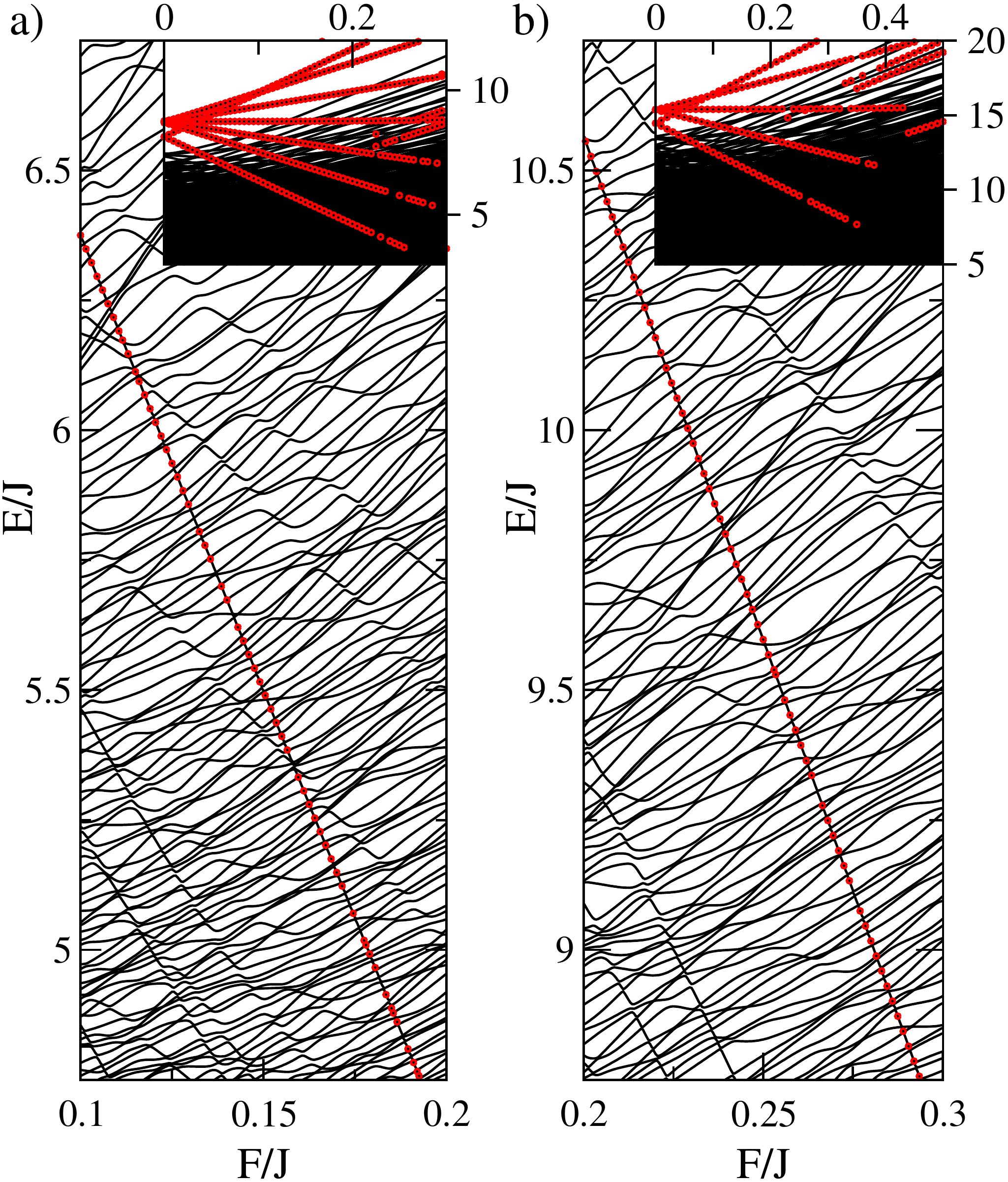}
\caption{
(Color online) Two spectra of the Bose-Hubbard-Hamiltonian (Eq.~(\ref{Eh})) as a function of the static tilt $F/J$ (cf. Fig.~\ref{Fspectrum}). The system parameters are a) interaction strength $U=0.5$, tunneling coupling $J=1$, $N=6$ particles, $M=7$ lattice sites, with Hilbert-space dimension ${\cal N}=1716$, and b) $U=0.3$, $J=1$, $N=10$, $M=5$, and ${\cal N}=1001$.
In both panels, the $M$ eigenstates with the largest IPR (see Eq.~(\ref{Eipr})) in the Fock number basis are plotted in red.
The insets show the spectrum on a larger scale.
}
\label{fig5}
\end{center}
\end{figure}

A few words regarding the dependence of our observations on the system parameters are in order.
As we will explain in detail in Section~5, the presence of solitonic states is not linked to a specific set of parameter values, but
we expect them quite generally to appear in a regime where the Bose-Hubbard spectrum can be considered primarily chaotic.
This expectation is corroborated by Fig.~\ref{fig5}, which shows the parametric evolution of two Bose-Hubbard lattices with a) $N=6$ particles in $M=7$ lattices sites,  and b) $N=10$ particles in $M=5$ lattice sites, and interaction values $U$ chosen in the chaotic regime \cite{Venzl11}.
In agreement with Fig.~\ref{Fspectrum}, we observe as many solitonic states (marked in red) as sites of the lattice -- $M=7$ and $M=5$ in Fig.~\ref{fig5} a) and b), respectively.
In the following, we fix $U=J=1$, $M=11$ sites, and $N=3$ bosons, as a typical case of well-developed quantum chaos in coexistence with solitonic states.

\begin{figure*}[t]
\begin{center}
\includegraphics[width=0.7\textwidth,type=pdf,ext=.pdf,read=.pdf]{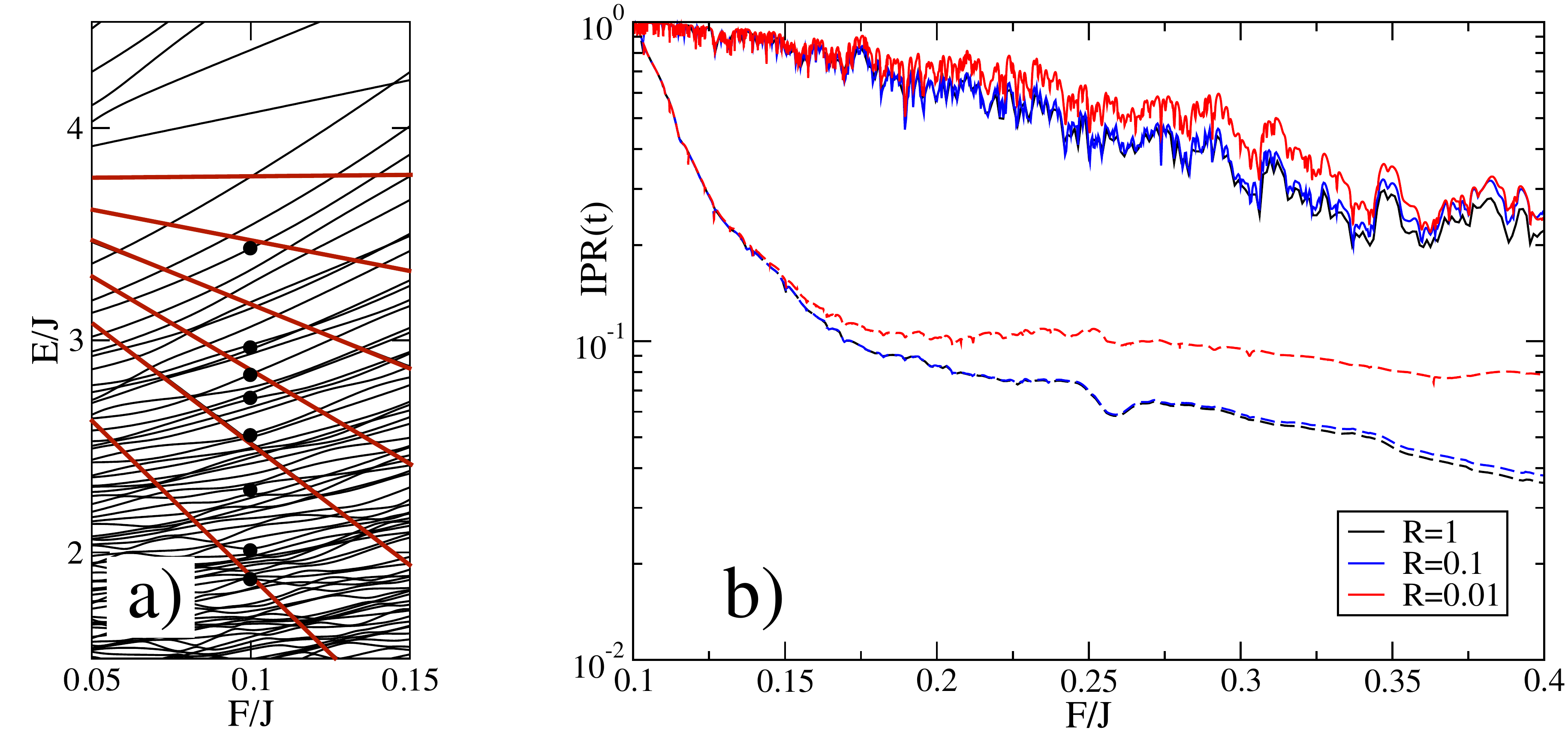}
\caption{(Color online) 
{\em Left:} Magnification of the spectrum where the energy levels for the initial states $\ket{\Psi}$ are marked. Black circles/red lines depict irregular/solitonic initial states.
{\em Right:} Dynamical evolution of the average IPR, \eq{Eipr}, in the instantaneous basis for various rates $R$, \eq{ER}, with $F_i/J=0.1$ and $F_f/J=0.4$. Solid/dashed lines represent solitonic/irregular initial states.}
\label{FIPRinst}
\end{center}
\end{figure*}

%
\section{Dynamical stability}
The solitonic states' invariance properties spelled out by the constant slope of their energy levels
 have an additional expression in the comparatively small avoided crossings with the irregular
states of the spectrum, as evident from \fig{Fspectrum}b).
This implies small coupling matrix elements between solitonic and irregular eigenstates, and therefore suggests an enhanced
stability of the solitonic states under variations of $F$.

To test this conjecture, we investigate the multi-particle dynamics under
a linear ramp of the tilt, $F(t)=F_{i} + R\, t$, from the initial value $F(0)=F_i$, to the final value $F(\Delta t)=F_f$.
That is, the tilt is varied with the slew rate
\begin{equation}
R=\frac{F_f-F_i}{\Delta t}\ ,
\label{ER}
\end{equation}
for both solitonic and irregular eigenstates $\ket{\Psi}$ of the BHH chosen from the same energy range in the {\em bulk}
of the spectrum, as shown in \fig{FIPRinst}a).
The stability of the initially prepared states is characterized in terms of the time-dependent IPR (see \eq{Eipr}),
for a given rate $R$.

\begin{figure*}[t]
\begin{center}
\includegraphics[width=0.9\textwidth,type=pdf,ext=.pdf,read=.pdf]{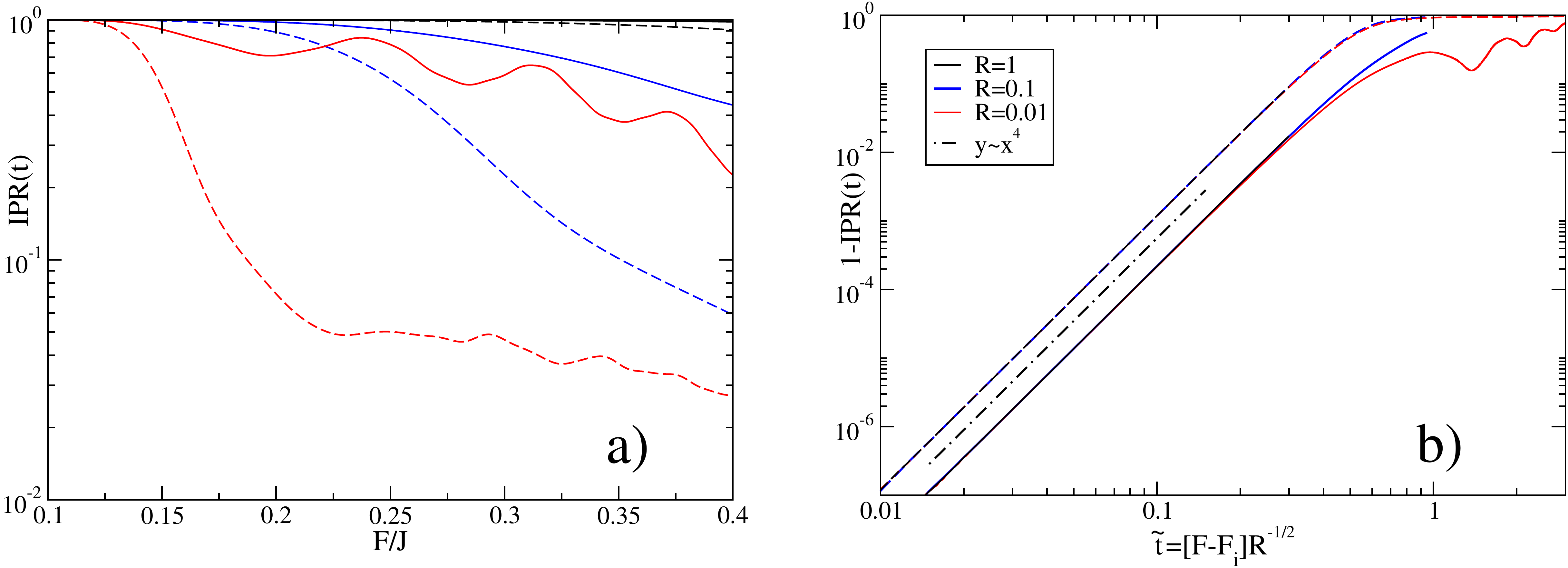}
\caption{(Color online) 
{\em Left:} Dynamical evolution of the average IPR, \eq{Eipr}, in the initial (fixed) eigenbasis for various rates $R$, \eq{ER}, with $F_i/J=0.1$ and $F_f/J=0.4$. Solid/dashed lines represent solitonic/irregular initial states.
{\em Right:} Drop of the IPR against the rescaled time $\tilde{t}=t\sqrt{R}$.
For short times, the curves for all three rates fall on top of each other, thereby forming the upper group of irregular states and the lower group of solitonic states. That is, both types of initial conditions exhibit excellent agreement with linear response theory -- which predicts \cite{CK00} a super-Gaussian decay for the survival probability $P(t)\sim\exp[-R^2t^4]$ (the dash-dotted line has slope four, and is drawn to guide the eye).
Note, however, the offset between irregular and solitonic states which amounts to about one order of magnitude.}
\label{FIPRfixed}
\end{center}
\end{figure*}

%
Let us first consider the IPR in the instantaneous basis,
\ie, in the Hamiltonian's time-dependent eigenbasis that follows the evolution of the static field $F(t)$.
It is shown in \fig{FIPRinst}b) and reveals the broadening of the initial state due to transitions to other modes.
We observe a strikingly different behavior for both initial conditions:
Whereas the IPR of the irregular states decays rapidly, the solitonic states' IPR decreases significantly slower. 
We also find that the decay rate observed in Fig. \ref{FIPRinst}b), is essentially independent of R, after an initial transient.
%
A complementary view on the stability can be obtained from the IPR in the initial (fixed) basis defined by $F_i$, see \fig{FIPRfixed}a).
For very large rates, $R=1$, the change in the tilt $F$ occurs on a time scale much faster than the internal (tunneling) dynamics on the lattice.
Hence, in this diabatic regime the system wave function cannot adapt to the changing potential.
Consequently, the IPR remains largely unaffected and a difference between solitonic and irregular states is hardly visible.
In contrast, as we reduce the ramp, we observe more and more pronounced drops of the IPR for both types of initial states.
However, the drop in the IPR of the irregular states is about one order of magnitude larger than that of the 
solitonic ones, what spells out the stability of the latter.


{We finally remark that the IPR's sensitive dependence on the slew rate $R$, present in the instantaneous basis shown in \fig{FIPRinst}b), can be understood from linear response theory. To this end, }we note that, for sufficiently short times, the inverse participation ratio ${\rm IPR(\ket{\Psi(t)})}=\sum_j|\bra{\Psi_j}\hat U(t)\ket{\Psi_0}|^4$ is essentially given by the
square of the survival probability $P(t)=|\bra{\Psi_0}\hat U(t)\ket{\Psi_0}|^2$ since $|\bra{\Psi_j}\hat U(t)\ket{\Psi_0}|^2\simeq 0$
for $j\neq 0$, where $\hat{U}(t)$ is the time-evolution operator.
For linearly driven chaotic systems, quantum linear response theory predicts a super-Gaussian decay $P(t)\sim\exp[-R^2t^4]$
\cite{CK00}, what suggests a scale invariance of $P(t)$ and hence of the IPR with respect to the scaled time $\tilde{t}=t\sqrt{R}$.
This is confirmed in \fig{FIPRfixed}b) where for solitonic {\em and} irregular initial states the same dependence is observed.
We stress, however, that between the two groups of curves there is an offset of about one order of magnitude, that is, the absolute decay is much weaker for the solitonic states 
\footnote{Note that the latter exhibit oscillations of $1-{\rm IPR}$ around $\tilde{t}=1$, which are likely to result from the
regularity of the integrable part of the spectrum.}.

%
\section{Generating mechanism} 
In the previous section, we numerically confirmed the dynamical stability of the solitonic states,
as suggested by their parametric level evolution.
We now turn to the discussion of the underlying mechanism \cite{VZOHMB10}.
The existence of eigenstates of a many-body system with all particles localized close to each other, despite
the presence of repulsive inter-particle interactions $U$, was experimentally first demonstrated for pairs of atoms \cite{Wink_etal06}. 
{Theoretical investigations of bound atom states as well as of bound quasi-particle states in nonlinear lattice systems, (see, e.g., \cite{Ovch69,SEG94,RM07,PSAF07,WHC08,WB09, JCS09}) also revealed the existence of bound states including more than two (quasi-) particles \cite{DESS04,VPS10,MSCE91,WGBS96,SWW96}.
Recently, also Bloch oscillations for initially localized states of interacting bosons were investigated \cite{KKHF10}.}
Such repulsively-bound many-particle states are formed as a consequence of the {\em energy mismatch} between
the on-site interaction energy and the maximal kinetic energy that can be realized in the lowest energy band in the lattice.
That is, they are energetically isolated from the remainder of the spectrum.

In contrast to that, for our presently solitonic states, which run through the {\em bulk} of the spectrum, also energetically allowed transitions (to the irregular states) are blocked, and, in addition, lead to an enhanced stability under perturbations.
This can be understood in terms of the transitions between the energy eigenstates that are induced by the tilt, as discussed in the following.
The {\em time-independent} part of the corresponding transition amplitude from some initial state $\ket{\Psi_i}$ to a final state
$\ket{\Psi_f}$ is given by the corresponding matrix element of the center-of-mass operator $\bra{\Psi_f}\sum_l\tilde{l}\hat n_l\ket{\Psi_i}$.
As shown by our analysis in \fig{FIPR} above, in case of an irregular state the atoms are distributed over the entire lattice while
the solitonic states are distinguished by the fact that most of the atoms occupy the same lattice site.
As a result, the solitonic states are approximate eigenstates of the center-of-mass operator and hence the matrix element
$\bra{\Psi_f}\sum_l\tilde{l}\hat n_l\ket{\Psi_i}$ with $\ket{\Psi_f}$ an irregular and $\ket{\Psi_i}$ a solitonic state
becomes very small, since both are system eigenstates {and thus, mutually orthogonal}.
More intuitively, a transition from a solitonic to an irregular state requires the redistribution of essentially all atoms over the entire lattice.
The tunneling of an atom over more than a single lattice site, as well as the simultaneous tunneling of more than one atom,
is, however, negligible for typical lattice-depths, since it corresponds to higher-order processes in the Bose-Hubbard Hamiltonian.
Whereas such processes can occur as resonance phenomena for large interaction strengths $U$ \cite{KKHF10},
they are negligible in the presently discussed regime of intermediate interactions.
Furthermore, in a tilted, infinitely long lattice (but with finite particle numbers), all states are localized \cite{KKHF10} and in that case, even analytic estimates of the localization volume of the states (and thus of the above transition matrix elements) can be obtained via composite-particle eigenvectors (see \cite{KKHF10}).

One might now wonder whether, similarly to the solitonic states, there are system eigenstates where (to a good approximation)
all atoms but one are located on a single site, and the remaining particle is localized on some other site.
Indeed, we find such states, that show comparable features as the solitonic states described so far, and they will be referred to as
{\em solitonic states of second order}.
For the case of three atoms, such states have also been analyzed in Ref.\cite{VPS10}.
For a transition from a solitonic state of first order to a solitonic state of second order only tunneling of a single atom over a single site
is necessary, and hence one expects significantly larger coupling as compared to irregular states.
Thus, in the simplest approximation one can treat two such states as a two-level system of (idealized) Fock states $\ket{\dots,0,N,0,\dots}$
and $\ket{\dots,1,N-1,0,\dots}$.
Assuming vanishing coupling $J=0$, their energy would coincide only at $F= U(N-1)$ which, for $U=1$, is far outside the regime in which solitonic
states exist (see Figs.\ref{Fspectrum} and \ref{FIPR}).
Once we take into account the tunneling, an avoided crossing of width $2J\sqrt{N}$ emerges between the two states and thus affects
the stability of the solitonic states for tilts beyond the threshold value $F_t\approx U(N-1)-J\sqrt{N}$.
For the parameter values used in Figs.\ref{Fspectrum} and \ref{FIPR}, this corresponds to $F_t\simeq 0.27$,
and this is indeed where the solitonic states start to dissolve.

If, for larger $N$, more particles are bound in a solitonic state, more have to undergo tunneling processes in order to transform into an irregular state.
From this point of view, the stability of the solitonic states is expected to be enhanced with increasing particle number at constant interaction strength $U$.
Yet, for an increasing number of particles located on a single site, three-body interactions that are not accounted for by the BHH might become non-negligible.
These three-body collisions may result in the formation of untrapped molecules leading to additional decay channels \cite{DGPS99} for solitonic states.
It is therefore crucial that the condition $UN=const.$ is met by the experiment, since then, three-body processes are not enhanced due to an effective reduction of the interaction strength $U$ as $N$ is increased.
The {\em same} condition is essential for the construction of a mean-field phase space for the Bose-Hubbard Hamiltonian, for large boson numbers and constant lattice length $M$ (see, \eg, \cite{HKG09,ELS85,FP03,HKG06,TWK09}):
The underlying principle is that under the requirement $UN=const.$ the mean-field dynamics is not changed as $N$ is increased, due to the scaling behavior of the BHH
\footnote{
We note that the parametric level dynamics presented in Figs.~\ref{Fspectrum} and \ref{fig5} obey this requirement, i.e., $UN=3$ in the three cases shown.
The mean-field dynamics is not the same, however, among lattices of {\it different} size $M$:
Given the fact that the lattice lengths $M$ were on purpose chosen differently, such as to demonstrate the general existence of solitonic states for various configurations, the respective phase spaces have different dimension of $2M$ \cite{ELS85} and are thus not identical.
}.
The phase space will, in general, be mixed regular and chaotic \cite{ELS85}.
The solitonic states should then be identifiable with regular islands that correspond to highly localized mean-field solutions as the ones that have been found, e.g., for three site systems (see, e.g., Fig.~6 of \cite{TWK09} and inset of Fig.~6 in \cite{Ng_etal09}).
Semiclassical arguments then guarantee the existence of quantum mechanical eigenstates which are localized within
these islands and are only weakly coupled to the states living on the chaotic sea.

\begin{figure*}[t]
\begin{center}
\includegraphics[width=.95\textwidth,type=pdf,ext=.pdf,read=.pdf,keepaspectratio,clip] {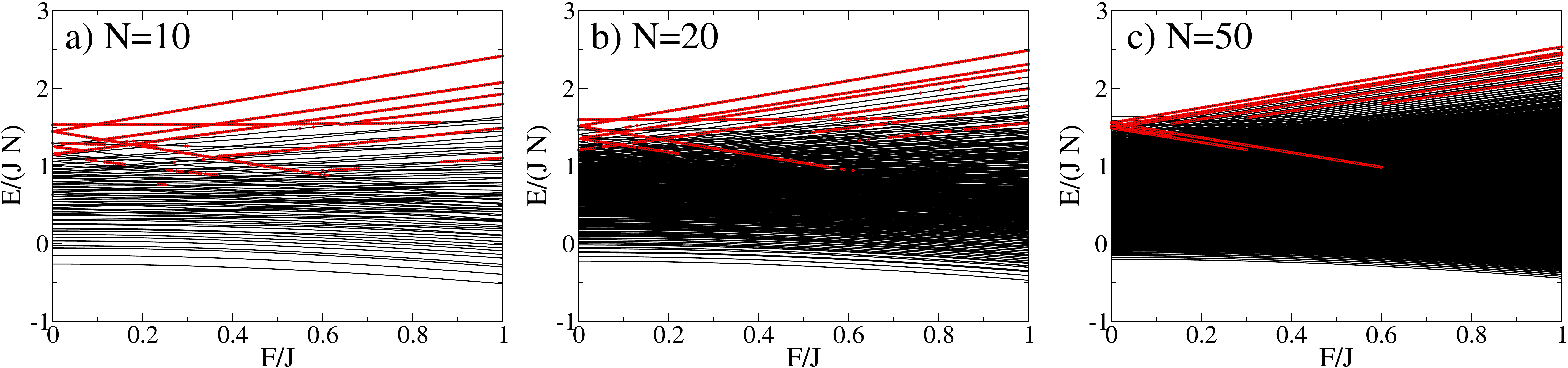}
\caption{
(Color online) The spectrum of the Bose-Hubbard Hamiltonian (Eq.~(\ref{Eh})) for fixed effective interaction $UN=3$ and three different particle numbers $N$, plotted as a function of the static tilt $F/J$ (cf. Fig.~\ref{Fspectrum}). For better comparability the energies are additionally scaled by $N$.
The system parameters are tunneling coupling $J=1$, $M=3$ lattice sites, and particle numbers (Hilbert-space dimension) a) $N=10$ $({\cal N}=66)$, b) $N=20$ $({\cal N}=231)$, and c) $N=50$ $({\cal N}=1326)$.
In all panels, only the six eigenstates with the largest IPR (see Eq.~(\ref{Eipr})) in the Fock number basis are plotted in red to keep the graphical presentation clear.
We note, though, that the number of solitonic states increases with $N$.
That level which crosses the bulk with negative slope and, in all three panels, is marked in red up to around $F/J\approx0.6$ corresponds to a solitonic state of first order.
}
\label{fig6}
\end{center}
\end{figure*}

While the analysis of the mean-field limit is beyond the scope of the present paper, we would like to shortly address some related aspects.
Due to the comparably small particle number $N\le10$ in the spectra presented in Figs.~\ref{Fspectrum} and \ref{fig5}, these systems are still well below the mean-field limit.
Instead, we show in Fig.~\ref{fig6} how the parametric level dynamics {\em itself} evolves as the particle number $N$ is increased, for fixed lattice size $M$ and constant effective interaction $UN=3$.
Owing to numerical manageability, we chose a three-site lattice ($M=3$), the smallest possible system for which chaos can be expected \cite{ELS85}, and increase the particle number from $N=10$ to $N=50$.
We first remark that in all cases solitonic states are indeed observed:
We highlighted the six levels with the highest IPR and note that in all three panels, the red level that crosses the bulk with negative slope and that is marked in red up to around $F/J\approx0.6$, corresponds to one of the three solitonic states of first order.
Second, qualitatively speaking, larger particle numbers result in a smaller effective Planck constant:
That is, for increasing $N$ and $UN=const.$, more quantum states share the available, $N$-independent phase space.
Thus, also more states can exist within the island, i.e., an increased number of solitonic states of higher order is expected.
This, as well, is confirmed by Fig.~\ref{fig6}: For $N=10$ particles, we find essentially one solitonic state with negative slope, corresponding to localization on the first lattice site, with negative single-particle on-site energy $F\tilde l=-F$, see below Eq.~(\ref{Eh}).
In contrast, for $N=20$, two such levels can be clearly distinguished while for $N=50$ already three such states are among the six levels with the highest IPR.
Finally, the detailed inspection of the spectra shown in Fig.~\ref{fig6} reveals that the critical tilt strength $F_t$, at which the solitonic state of highest order disappears, displays a weak dependence on the particle number $N$.
Namely, for $N=10$ $(N=50)$ we determine $F_t$ to be approximately $F/J\approx0.65$ $(F/J\approx0.75)$.
This indicates that for increasing $N$, the criterion for $F_t$ should be based on the following semiclassical consideration instead of the simplified picture given above:
As the tilt $F$ is increased, the stable island (associated with localization on the corresponding lattice site) shrinks and, thus, fewer and fewer solitonic states can reside on it.
Finally, at a certain value of the tilt $F$, given by the ($N$-independent) mean-field equations, the island vanishes and this determines the critical value $F_t$ at which the solitonic state of highest order disappears
\footnote{
For a two-site Bose-Hubbard system a similar strategy was successfully applied in Ref.~\cite{SCHKC09} (see also references therein), and a tilted chaotic three-mode system has been studied in the mean-field limit in Refs.~\cite{Wang06,GKW06}.}.

With respect to the experimental viability of the investigated setup, it is fair to say that not only the lattice geometry is under exquisite control by the experimentalists, but also interaction strength and particle detection:
In typical optical lattice-based experiments, the ratio $U/J$ can be sensitively adjusted by a variation of the lattice depth \cite{JBCGZ98,BDZ08}.
Depending on the atomic species, $U$ can additionally be controlled, independently of the coupling tunneling $J$, via Feshbach resonances \cite{IASMSK98, CGJT10}.
For example, in experiments with Cs atoms, the scattering length could be tuned from small negative to very large positive values, over several orders of magnitude \cite{GHMDRN08}.
In the same setup, magnetic levitation was rapidly switched off within about hundred $\mu$s and was used to compensate for gravitation.
Presumably, this levitation can as well be tuned to a nonzero value as to adjust the strength of the gravitational forces, what would amount to controlling the lattice tilt $F$.
Finally, we note that the detection efficiency has reached the single-atom level \cite{Bakr09,Sher10}, what permits a very precise counting of occupation numbers. Thus, the experimental detection of solitonic states should be within reach.

%
\section{Conclusion}
In summary, we have studied the dynamical evolution of ultracold bosons confined on a one-dimensional optical lattice, which is subject to a tilt.
Our present work was focused on the parameter regime in which kinetic, inter-particle interaction, and on-site energy are balanced.
In this regime, where the corresponding Bose-Hubbard Hamiltonian with an additional tilt term can be characterized as being primarily chaotic in a spectral sense \cite{BK03,KB03,Venzl11}, we have identified regular structures in the parametric level evolution.
Associated with these structures are the solitonic states, which are distinguished by a strong localization on the lattice as well as in the Fock space, which barely changes with the tilt strength.
Unlike other studies on energetically isolated bound states \cite{Wink_etal06,DESS04,VPS10}, this property results from the solitonic states' weak coupling to the bulk of energy levels \cite{VZOHMB10}.

Based on the inverse participation ratio, evaluated in the fixed as well as in the adiabatic basis of the Bose-Hubbard Hamiltonian, we compared the dynamical stability of these solitonic states to an ensemble of neighboring states in the bulk of the spectrum, as the tilt was ramped with different slew rates.
We observed a drastically increased robustness of the solitonic states, spelled out by a significantly larger inverse participation ratio at the end of the dynamical evolution.

The present work belongs to that branch of ultracold atom physics, which is concerned with comparatively small particle numbers.
This direction has recently seen substantial experimental advances, not at last in the resolution of imaging techniques \cite{Bakr09,Sher10} that would allow to directly identify the solitonic states.
Their remarkable dynamical robustness makes them excellent candidates, e.g., for the preparation of stable quantum few-body states, in a 
parameter regime where, due to the generic presence of spectral chaos, a substantial portion of the eigenstates sensitively depends on  system parameters such as the tilt.
On the other hand, provided that coherent superpositions of solitonic states can be prepared, reasonably stable tunnel dynamics of more than one boson \cite{KKHF10} would be realizable for relatively small interaction values.
Finally, in the light of the mean-field limit discussed in the previous section, the crossover in stability of these states from the few- to the many-body regime might be explored. 

%
\ack

We acknowledge financial support of the Deutsche Forschungsgemeinschaft DFG via the {\em Research Unit 760}, and through a personal grant (F.M.).
B.O. gratefully acknowledges financial support of the European Science Foundation within the
QUDEDIS program and of the Polish Government (scientific funds 2008-2011).

\ \\ \ \\
\bibliographystyle{unsrt}

\bibliography{soliton}

\begin{thebibliography}{10}

\bibitem{DGPS99}
F.~Dalfovo, S.~Giorgini, L.~P. Pitaevskii, and S.~Stringari.
\newblock {T}heory of {B}ose-{E}instein condensation in trapped gases.
\newblock {\em Rev. Mod. Phys.}, 71:463, 1999.

\bibitem{Fert05}
C.~D. Fertig, K.~M. O'Hara, J.~H. Huckans, S.~L. Rolston, W.~D. Phillips, and
  J.~V. Porto.
\newblock {Strongly Inhibited Transport of a Degenerate 1D Bose Gas in a
  Lattice}.
\newblock {\em Phys. Rev. Lett.}, 94:120403, 2005.

\bibitem{PMKB06}
A.~V. Ponomarev, J.~Madronero, A.~R. Kolovsky, and A.~Buchleitner.
\newblock {A}tomic {C}urrent across an {O}ptical {L}attice.
\newblock {\em Phys. Rev. Lett.}, 96:050404, 2006.

\bibitem{MO06}
O.~Morsch and M.~Oberthaler.
\newblock {D}ynamics of {B}ose-{E}instein condensates in optical lattices.
\newblock {\em Rev. Mod. Phys.}, 78:179, 2006.

\bibitem{BDZ08}
I.~Bloch, J.~Dalibard, and W.~Zwerger.
\newblock Many-body physics with ultracold gases.
\newblock {\em Rev. Mod. Phys.}, 80:885, 2008.

\bibitem{FFS89}
S.~de~Filippo, M.~Fusco~Girard, and M.~Salerno.
\newblock {A}voided crossing and nearest-neighbour level spacings for the
  quantum {DST} equation.
\newblock {\em Nonlinearity}, 2:477, 1989.

\bibitem{Chef96}
A.~Chefles.
\newblock {N}earest-{N}eighbour level spacings for the non-periodic discrete
  {S}chr\"odinger equation.
\newblock {\em J. Phys. A.}, 29:4515, 1996.

\bibitem{KB04}
A.~R. Kolovsky and A.~Buchleitner.
\newblock {Q}uantum chaos in the {B}ose-{H}ubbard model.
\newblock {\em Europhys. Lett.}, 68:632, 2004.

\bibitem{HKG09}
M.~Hiller, T.~Kottos, and T.~Geisel.
\newblock {W}ave-packet dynamics in energy space of a chaotic trimeric
  {B}ose-{H}ubbard system.
\newblock {\em Phys. Rev. A}, 79:023621, 2009.

\bibitem{Chirikov79}
B.~V. Chirikov.
\newblock A universal instability of many-dimensional oscillator systems.
\newblock {\em Phys. Rep.}, 52:263, 1979.

\bibitem{Buchleitner2002409}
A.~Buchleitner, D.~Delande, and J.~Zakrzewski.
\newblock Non-dispersive wave packets in periodically driven quantum systems.
\newblock {\em Physics Reports}, 368:409, 2002.

\bibitem{Wink_etal06}
K.~Winkler, G.~Thalhammer, F.~Lang, R.~Grimm, J.~H. Denschlag, A.~J. Daley,
  A.~Kantian, H.~P. B\"uchler, and P.~Zoller.
\newblock Repulsively bound atom pairs in an optical lattice.
\newblock {\em Nature}, 441:853, 2006.

\bibitem{DESS04}
J.~Dorignac, J.~C. Eilbeck, M.~Salerno, and A.~C. Scott.
\newblock {Quantum Signatures of Breather-Breather Interactions}.
\newblock {\em Phys. Rev. Lett.}, 93:025504, 2004.

\bibitem{VPS10}
M.~Valiente, D.~Petrosyan, and A.~Saenz.
\newblock Three-body bound states in a lattice.
\newblock {\em Phys. Rev. A}, 81:011601(R), 2010.

\bibitem{Haa00}
F.~Haake.
\newblock {\em {Q}uantum {S}ignatures of {C}haos}.
\newblock Springer-Verlag, Berlin, Heidelberg, New York, second edition, 2000.

\bibitem{JBCGZ98}
D.~Jaksch, C.~Bruder, J.~I. Cirac, C.~W. Gardiner, and P.~Zoller.
\newblock {C}old {B}osonic {A}toms in {O}ptical {L}attices.
\newblock {\em Phys. Rev. Lett.}, 81:3108, 1998.

\bibitem{IASMSK98}
S.~Inouye, M.~R. Andrews, J.~Stenger, H.-J. Miesner, D.~M. Stamper-Kurn, and
  W.~Ketterle.
\newblock {O}bservation of {F}eshbach resonances in a {B}ose-{E}instein
  condensate.
\newblock {\em Nature}, 392:151, 1998.

\bibitem{CGJT10}
C.~Chin, R.~Grimm, P.~Julienne, and E.~Tiesinga.
\newblock Feshbach resonances in ultracold gases.
\newblock {\em Rev. Mod. Phys.}, 82:1225, 2010.

\bibitem{ELS85}
J.~C. Eilbeck, P.~S. Lomdahl, and A.~C. Scott.
\newblock {T}he discrete self-trapping equation.
\newblock {\em Physica D}, 16:318, 1985.

\bibitem{SFGS97}
A.~Smerzi, S.~Fantoni, S.~Giovanazzi, and S.R. Shenoy.
\newblock {Q}uantum {C}oherent {A}tomic {T}unneling between {T}wo {T}rapped
  {B}ose-{E}instein {C}ondensates.
\newblock {\em Phys. Rev. Lett.}, 79:4950, 1997.

\bibitem{VA00}
A.~Vardi and J.~R. Anglin.
\newblock {B}ose-{E}instein {C}ondensates beyond {M}ean {F}ield {T}heory:
  {Q}uantum {B}ackreaction as {D}ecoherence.
\newblock {\em Phys. Rev. Lett.}, 86:568, 2000.

\bibitem{AV01}
J.~R. Anglin and A.~Vardi.
\newblock {D}ynamics of a two-mode {B}ose-{E}instein condensate beyond
  mean-field theory.
\newblock {\em Phys. Rev. A}, 64:013605, 2001.

\bibitem{SSAC09}
K.~Sakmann, A.~I. Streltsov, O.~E. Alon, and L.~S. Cederbaum.
\newblock {Exact Quantum Dynamics of a Bosonic Josephson Junction}.
\newblock {\em Phys. Rev. Lett.}, 103:220601, 2009.

\bibitem{SSAC10}
K.~Sakmann, A.~I. Streltsov, O.~E. Alon, and L.~S. Cederbaum.
\newblock {Quantum dynamics of attractive versus repulsive bosonic Josephson
  junctions: Bose-Hubbard and full-Hamiltonian results}.
\newblock {\em Phys. Rev. A}, 82:013620, 2010.

\bibitem{Chuc10}
M.~Chuchem, K.~Smith-Mannschott, M.~Hiller, T.~Kottos, A.~Vardi, and D.~Cohen.
\newblock Quantum dynamics in the bosonic {J}osephson junction.
\newblock {\em Phys. Rev. A}, 82:053617, 2010.

\bibitem{KB03}
A.~R. Kolovsky and A.~Buchleitner.
\newblock {F}loquet-{B}loch operator for the {B}ose-{H}ubbard model with static
  field.
\newblock {\em Phys. Rev. E}, 68:056213, 2003.

\bibitem{Venzl11}
Hannah Venzl.
\newblock {\em Ultracold bosons in tilted optical lattices --- impact of
  spectral statistics on simulability, stability, and dynamics}.
\newblock PhD thesis, Albert-Ludwigs Universit\"at Freiburg, Germany, 2011,
  {\tt http://www.freidok.uni-freiburg.de/volltexte/8126/}.

\bibitem{Feyn39}
R.~P. Feynman.
\newblock Forces in {M}olecules.
\newblock {\em Phys. Rev.}, 56:340, 1939.

\bibitem{ZBD97}
J.~Zakrzewski, A.~Buchleitner, and D.~Delande.
\newblock {N}ondispersive wave packets as solitonic solutions of level
  dynamics.
\newblock {\em Z. Phys. B}, 103:115, 1997.

\bibitem{GGK02}
M.~Gl\"uck, A.R. Kolovsky, and H.~J. Korsch.
\newblock {Wannier--Stark resonances in optical and semiconductor
  superlattices}.
\newblock {\em Phys. Rep.}, 366:103, 2002.

\bibitem{CK00}
D.~Cohen and T.~Kottos.
\newblock {Q}uantum-{M}echanical {N}onperturbative {R}esponse of {D}riven
  {C}haotic {M}esoscopic {S}ystems.
\newblock {\em Phys. Rev. Lett.}, 85:4839, 2000.

\bibitem{VZOHMB10}
H.~Venzl, T.~Zech, B.~Ole\'s, M.~Hiller, F.~Mintert, and A.~Buchleitner.
\newblock {S}olitonic eigenstates of the chaotic {B}ose--{H}ubbard
  {H}amiltonian.
\newblock {\em Appl. Phys. B: Lasers Opt.}, 98:647, 2010.

\bibitem{Ovch69}
A.A. Ovchinnikov.
\newblock Localized long-lived vibrational states in molecular crystals.
\newblock {\em Zh. Eksp. Teor. Fiz. / Soviet Phys. JETP}, 57/30:263/147, 1969.

\bibitem{SEG94}
A.~C. Scott, J.~C. Eilbeck, and H.~Gilh{\o}j.
\newblock Quantum lattice solitons.
\newblock {\em Physica D}, 78:194, 1994.

\bibitem{RM07}
R.~Piil and K.~M\o{}lmer.
\newblock Tunneling couplings in discrete lattices, single-particle band
  structure, and eigenstates of interacting atom pairs.
\newblock {\em Phys. Rev. A}, 76:023607, 2007.

\bibitem{PSAF07}
D.~Petrosyan, B.~Schmidt, J.~R. Anglin, and M.~Fleischhauer.
\newblock Quantum liquid of repulsively bound pairs of particles in a lattice.
\newblock {\em Phys. Rev. A}, 76:033606, 2007.

\bibitem{WHC08}
L.~Wang, Y.~Hao, and S.~Chen.
\newblock {Quantum dynamics of repulsively bound atom pairs in the Bose-Hubbard
  model}.
\newblock {\em Eur. Phys. J. D}, 48:229, 2008.

\bibitem{WB09}
Ch. Weiss and H.-P. Breuer.
\newblock Photon-assisted tunneling in optical lattices: Ballistic transport of
  interacting boson pairs.
\newblock {\em Phys. Rev. A}, 79:023608, 2009.

\bibitem{JCS09}
L.~Jin, B.~Chen, and Z.~Song.
\newblock {Coherent shift of localized bound pairs in the Bose-Hubbard model}.
\newblock {\em Phys. Rev. A}, 79:032108, 2009.

\bibitem{MSCE91}
P.D. Miller, A.C. Scott, J.~Carr, and J.C. Eilbeck.
\newblock {Binding energies for discrete nonlinear Schr\"odinger equations}.
\newblock {\em Phys. Scr.}, 44:509, 1991.

\bibitem{WGBS96}
W.Z. Wang, J.~Tinka Gammel, A.R. Bishop, and M.I. Salkola.
\newblock Quantum breathers in a nonlinear lattice.
\newblock {\em Phys. Rev. Lett.}, 76:3598, 1996.

\bibitem{SWW96}
S.A. Schofield, R.E. Wyatt, and P.G. Wolynes.
\newblock {Computational study of many-dimensional quantum vibrational energy
  redistribution. i. Statistics of the survival probability}.
\newblock {\em J. Chem. Phys.}, 105:940, 1996.

\bibitem{KKHF10}
R.~Khomeriki, D.~O. Krimer, M.~Haque, and S.~Flach.
\newblock Interaction-induced fractional {B}loch and tunneling oscillations.
\newblock {\em Phys. Rev. A}, 81:065601, 2010.

\bibitem{FP03}
R.~Franzosi and V.~Penna.
\newblock {C}haotic behavior, collective modes, and self-trapping in the
  dynamics of three coupled {B}ose-{E}instein condensates.
\newblock {\em Phys. Rev. E}, 67:046227, 2003.

\bibitem{HKG06}
M.~Hiller, T.~Kottos, and T.~Geisel.
\newblock {C}omplexity in parametric {B}ose-{H}ubbard {H}amiltonians and
  structural analysis of eigenstates.
\newblock {\em Phys. Rev. A}, 73:061604(R), 2006.

\bibitem{TWK09}
F~Trimborn, D~Witthaut, and H.~J. Korsch.
\newblock Beyond mean-field dynamics of small {B}ose-{H}ubbard systems based on
  the number-conserving phase space approach.
\newblock {\em Phys. Rev. A}, 79:013608, 2009.

\bibitem{Ng_etal09}
G.~S. Ng, H.~Hennig, R.~Fleischmann, T.~Kottos, and T.~Geisel.
\newblock {Avalanches of Bose--Einstein condensates in leaking optical
  lattices}.
\newblock {\em New J. Phys.}, 11:073045, 2009.

\bibitem{SCHKC09}
K.~Smith-Mannschott, M.~Chuchem, M.~Hiller, T.~Kottos, and D.~Cohen.
\newblock {O}ccupation {S}tatistics of a {B}ose-{E}instein {C}ondensate for a
  {D}riven {L}andau-{Z}ener {C}rossing.
\newblock {\em Phys. Rev. Lett.}, 102:230401, 2009.

\bibitem{Wang06}
G.-F. Wang, D.-F. Ye, L.-B. Fu, X.-Z. Chen, and J.~Liu.
\newblock {Landau-Zener tunneling in a nonlinear three-level system}.
\newblock {\em Phys. Rev. A}, 74:033414, 2006.

\bibitem{GKW06}
E.~M. Graefe, H.~J. Korsch, and D.~Witthaut.
\newblock Mean-field dynamics of a {B}ose-{E}instein condensate in a
  time-dependent triple-well trap: {N}onlinear eigenstates, {L}andau-{Z}ener
  models, and stimulated {R}aman adiabatic passage.
\newblock {\em Phys. Rev. A}, 73:013617, 2006.

\bibitem{GHMDRN08}
M.~Gustavsson, E.~Haller, M.~J. Mark, J.~G. Danzl, G.~Rojas-Kopeinig, and H.-C.
  N{\"a}gerl.
\newblock Control of {I}nteraction-{I}nduced {D}ephasing of {B}loch
  {O}scillations.
\newblock {\em Phys. Rev. Lett.}, 100:080404, 2008.

\bibitem{Bakr09}
W.~S. Bakr, J.~I. Gillen, A.~Peng, S.~Folling, and M.~Greiner.
\newblock A quantum gas microscope for detecting single atoms in a
  {H}ubbard-regime optical lattice.
\newblock {\em Nature}, 462:74, 2009.

\bibitem{Sher10}
J.~F. Sherson, C.~Weitenberg, M.~Endres, M.~Cheneau, I.~Bloch, and S.~Kuhr.
\newblock {Single-atom-resolved fluorescence imaging of an atomic Mott
  insulator}.
\newblock {\em Nature}, 467:68, 2010.

\bibitem{BK03}
A.~Buchleitner and A.~R. Kolovsky.
\newblock {I}nteraction-induced decoherence of atomic {B}loch oscillations.
\newblock {\em Phys. Rev. Lett.}, 91:253002, 2003.

\end{thebibliography}

\end{document}